# Assessment of Technical Efficiency in the Moroccan Public Hospital Network: Using the DEA Method.

Er-Rays Youssef [1*][0000-0002-6691-9226], M'dioud Meriem [2**][0000-0002-7855-3495]

[1] Ibn Tofail University, Economics and Management Faculty (FEG), National School of Business and Management (ENCG), Research Laboratory in Organizational Management Sciences, Kenitra, Morocco
[2] Laboratory Engineering Sciences ENSA, Ibn Tofail University, Kenitra, Morocco
*raysyoussef@gmail.com

** meriemmdioud@gmail.com

**Abstract.**
**Background:** The public hospital network in Morocco plays a crucial role in providing healthcare services. However, this network faces challenges in terms of technical efficiency in healthcare management.
**Objectives:** This study aimed to assess the technical efficiency of the public hospital network in Morocco.
**Methods:** This article compares the efficiency of 77 public hospital networks from 2017 to 2020. Data were collected from the Directorate of Planning and Financial Resources (DPFR) of the Health Ministry Marocco. The Data Envelopment Analysis (DEA) method was employed, using three inputs (Hospital, Physician and Paramedical) and four outputs (Functional capacity, Hospitalization days and Admission). Additionally, the Malmquist index (MI) is utilized to analyse the factors of production, and peer modelling is incorporated to address hospital inefficiency.
**Results:** The average technical efficiency of public hospital networks under the CRS hypothesis from 2017 to 2020 is 0.697 (71% of DMUs have a score lower than 1), indicating that these networks need to minimize their inputs by approximately 30.3%. The Malmquist index reveals a decline in productivity gain from 2017/2018 (score of 0.980) followed by improvement in 2018/2019 (score of 1.163). In terms of peer modelling, 72.7% of the DMUs should emulate the most effective DMUs beginning in 2020, whereas the lowest score was observed in 2019**.**
**Conclusion:** These findings highlight the need for the public hospital network in Morocco to enhance the effective and efficient utilization of inputs, such as the number of hospitals and medical and paramedical staff, to produce the same outputs, including the number of surgical procedures, hospitalization days, admissions, and functional capacity.

## 1 Background

Public health hospitals play a crucial role in promoting health, ensuring access to care, and providing curative services for high-risk diseases (WHO 2000). Management and governance bodies that the state is working to create include (Ministry of Health 2022): - A higher health authority, responsible primarily for the technical oversight of basic compulsory health insurance, evaluating the quality of health institutions' services, and providing input on public health policies. - Territorial health groups, structured as public institutions, tasked with implementing state health policies at the regional level. Each group encompasses all public sector health institutions within its territorial jurisdiction. And - Two public institutions: one overseeing medicines and health products, and the other managing blood and its derivatives (Chapter Ten, Article 32) .The organisation of Moroccan hospital care provision is governed by the complementarity of the current hospital regulations, namely the reforms and restructuring implemented since 1983 (Ministry of Health 1983). The hospital reform of 1993, initiated through ministerial dissemination, was followed by the Hospital Organisation Decree of 2007 (Decree No. 2-06-656) (Ministry of Health 2007) , the internal regulations of hospitals in 2010 (Order 456-11, dated July 6, 2010) (Ministry of Health 2010) and framework law No. 06-22 promulgated by Dahir No. 1-22-77 of 14 Joumada I 1444 (December 9, 2022) relating to the national health system (Ministry of Health 2022).
Hospitals are organised into four types based on the range of their services and the nature of their equipment (general hospitals and specialised hospitals) (Ministry of Health 1983, 2007, 2010), as well as the budgetary management mode, which includes State Managed Autonomous Services (SEGMA) lacking legal personality but having financial autonomy, such as CHP, CHR, and CHIR. In contrast, Public Hospital Establishments (EPH) have legal personality and financial autonomy, such as university hospital centres (CHU). They vary according to the scope of action and level of provision (provincial hospital (CHP), regional (CHR), and interregional (CHIR)), and the size of the establishments (less than 120 beds, 120 to 240 beds, and more than 240 beds).

According to the 2010 reform (Ministry of Health 2010), the hospital is organised into three management poles, namely the Medical Affairs Pole (PAM), Nursing Care (PSI), and Administrative Affairs (PAA). Six planning, coordination, consultation, and support bodies include the Establishment Committee (CE), Monitoring and Evaluation Committee (CSE), Nosocomial Infection Control Committee (CLIN), as well as the Council of Physicians, Dentists, and Pharmacists (CMDP), and Nurses (CII). Two new services comprise the Admission and Reception Service (SAA) and the Hospital Pharmacy Service (SPH).

Health systems are responsible for promoting global health by ensuring accountability, efficiency, justice, and responsiveness to individuals of all ages and socioeconomic strata (Er-Rays et Ait-Lemqeddem 2021b; Smith et al. 2010). As a result, the purpose of this study was to estimate the optimal combination of healthcare hospital variables for improving the quality of healthcare provided to Moroccan patients.

However, hospitals in Morocco face various dysfunctions, including financial constraints, performance issues in healthcare, and the underutilization of material and human resources, which can negatively impact healthcare production (Er-Rays 2021; Er-Rays et Ait Lemqeddem 2020; Er-Rays et Ait-Lemqeddem 2021b).

This study contributes to the literature review and evaluation of the performance of public hospital healthcare in Morocco. The booklet has been organized into five chapters. Beginning with a literature review, we proceed to the techniques, findings, discussion, and conclusion and recommendations. The objective of this study was to evaluate the technical efficiency (TE) of seventy-seven Moroccan hospitals between 2017 and 2020, which is defined by their inherent complexity. Data envelopment analysis (DEA), the Malmquist index (MI), and modelling peers (MPs) were used.

## 2  Literature review

Analysing the performance of healthcare hospitals is essential for ensuring the sustainability of the population (Er-Rays et Ait-Lemqeddem 2021a). As a result, monitoring efficiency is an important technique for determining performance (Vrijens et al. 2014). It also entails developing a conceptual framework for modelling the system's components and identifying performance metrics. Husseiny measures the performance of healthcare systems, making it easier to identify information collection gaps (El Husseiny 2022).

Data envelopment analysis (DEA) is a method used to measure efficiency. DEA is a nonparametric technique for measuring the effectiveness of various social and economic systems (Cetin et Bahce 2016; Charnes, Cooper, et Rhodes 1978; Jung et al. 2023). It was first introduced by Farrel in 1957 (Farrell 1957) and was subsequently developed by various researchers, including (Charnes et al. 1978; Debreu 1951; Färe et al. 1994; Färe, Grosskopf, et Norris 1997; Koopmans 1951).

Numerous studies have examined hospital efficiency using the DEA model. (El Husseiny 2022) analysed healthcare systems for Arab nations (El Husseiny 2022). Furthermore, DEA has been frequently employed in a number of African nations to assess the technical effectiveness of hospitals (Top, Konca, et Sapaz 2020) (Ali, Debela, et Bamud 2017). Several related studies have been undertaken in Morocco (Er-Rays et Ait Lemqeddem 2020), Benin (Asbu et al. 2003), Eastern Ethiopia (Ali et al. 2017), Botswana (Tlotlego et al. 2010), Burkina Faso (Marschall et Flessa 2009), South Africa (Kirigia 2001) and Eritrea (Kirigia et Asbu 2013). Most of these studies focused on the first stage of efficiency analysis, which considers either curative treatment supplied by hospitals or preventative care provided by primary healthcare organizations. In this work, we focus on the first step of the analysis, which includes both types of facilities and examines the causes of technical efficiency (or inefficiency) using panel data.

Each of these works develops its research according to the data available and the perceptions of the authors concerning the choice of input and output variables. Patient quality of care, in terms of safety and efficiency, has significantly improved population health, as indicated by increased life expectancy and reduced infant mortality rates. Charnes et al. proposed an input-oriented DEA model assuming constant returns to scale (CRS) (Charnes et al. 1978), while other authors, such as Banker et al., introduced the concept of variable returns to scale (VRS) (Banker, Charnes, et Cooper 1984).

The combination of DEA with the Malmquist index (MI) enables the analysis of total production efficiency over multiple periods and provides a measure of total factor productivity that complements the DEA method. (Malmquist 1953) presented the foundation for this index, which was further developed into a productivity index by (Caves, Christensen, et Diewert 1982). (Färe et al. 1994) introduced the MI framework in the DEA literature. One of the advantages of the DEA method is its ability to provide benchmarking by estimating peer models for each inefficient healthcare facility and identifying efficient facilities that closely approach the efficiency frontier (Coelli 1996; Huguenin 2013).

## 3 Research methodology

### 3.1 Data and Variables

This study examines the performance of provincial or prefectural public hospital networks in Morocco and the impact of these networks on healthcare provision. Utilizing the DEA method and the Malmquist Index, this study benchmarks seventy-seven hospitals in Morocco's public health networks. The Ministry of Health's data from 2017–2020 were used for analysis (Health 2017, 2018, 2019, 2020); these data were obtained from the Directorate of Planning and Financial Resources (DPFR), produced by the Planning and Studies Division, Health Studies, and Information
Service, which is a primary source of data used. We examined the networks in each province or prefecture.

The Moroccan public hospital has undergone numerous administrative and care reorganizations as part of hospital reform, encompassing care, training, research, and social and population responsibilities. The hospital offering now consists of 61 provincial hospital centers, 12 regional hospital centers, and 06 inter-regional university hospital centers, totaling around 162 hospitals. However, rapid advancements in medical technology, digitalization, societal changes, and disease management techniques pose significant challenges that hospitals must adapt to quickly. Hospital supply increased from 155 hospitals in 2015 to 162 in 2020, a 4.32% increase from 2015 to 2020. However, this supply does not correspond to the increase in population and urbanization. These services are distributed across the seventy-seven provinces or prefectures (PPs) within twelve regional directorates. Seventy-seven public hospital networks were selected for further analysis to identify three inputs and four outputs (Table 1).

Table 1. Inputs and Outputs Variables (PHN: 2017-2020)

| variables | | Justification |
|---|---|---|
| $X_1$ | Hospital | Hospitals are the principal institutions that offer medical treatment to people. They serve as the focal point for a variety of healthcare activities, such as diagnosis, treatment, surgery, and rehabilitation. Improving hospital efficiency has a direct influence on patient flow, resource use, and overall service delivery. |
| $X_2$ | Physician | Physicians have a critical role in providing medical services such as diagnosis, treatment planning, and patient care. Their efficiency has a direct influence on the quality and speed of healthcare delivery in hospitals. |
| $X_3$ | Paramedical | Paramedical personnel, which includes nurses, technicians, and allied healthcare professionals, offer important support services to ensure that hospital operations run smoothly. Improving their efficiency has a substantial influence on patient care outcomes and overall hospital performance. |
| $Y_1$ | Surgical interventions | Surgical operations are an essential component of hospital treatment, frequently necessitating large resources and coordination. By improving surgical treatments, hospitals may improve patient outcomes, minimize surgery wait times, and increase overall operating room efficiency. |
| $Y_2$ | Functional capacity | Functional capacity refers to the hospital's ability to effectively use its resources, such as personnel, equipment, and facilities, to satisfy patient demands. Improving functional capacity may result in improved resource allocation, fewer bottlenecks, and more overall efficiency. |
| $Y_3$ | Hospitalization days | Hospitalization days reflect the amount of time patients spend in the hospital, which has a direct influence on resource use, bed occupancy rates, and overall efficiency. Reducing hospitalization days may result in financial savings, improved patient flow, and easier access to hospital services. |
| $Y_4$ | Admissions | Admissions show the number of patients entering the hospital, which may strain resources and reduce overall efficiency. By streamlining admissions procedures, hospitals may better control patient flow, minimize wait times, and increase access to treatment. |

### 3.1 Objective function of technical efficiency

The calculation of technical efficiency relies on the DEA method, which involves the virtual combination of ratios between all outputs and inputs of each DMU. This is achieved by utilizing the CRS and VRS models with input orientation [27].

The mathematical relationship between these two models is expressed as TE CRS = TE VRS (SE), where SE represents scale efficiency. This relationship indicates that the technical efficiency of an organization, under constant returns to scale (CRS), can be separated into two distinct components: pure technical efficiency and scale efficiency (SE).

The challenge faced by each decision-making unit (DMU) is to determine the optimal weights by solving a mathematical programming problem using linear programming duality. It can be assumed that healthcare establishments are categorized based on their optimal production scale. However, constraints such as the structure of healthcare establishments, financial limitations, and organizational issues have led to deviations from this optimal scale.

The objective function of the DEA method draws inspiration from previous works by (Er-Rays 2021; Er-Rays et Ait Lemqeddem 2020; Rays et Lemqeddem 2020), as well as from various other model specifications commonly employed in the healthcare domain.

**Measurement of efficiency: DEA.**

We can consider the total number of PHN public hospital networks in each province or prefecture. Each establishment uses a combination of U inputs to produce U outputs.

$A_{hch=}(a_{1phn}, a_{2phn}, \ldots, a_{Uphn})$: observed input vectors of the i-th public hospital network (PHN), and
$B_{hch=}(b_{1phn}, b_{2phn}, \ldots, b_{U'phn})$: observed output vectors of the i-th public hospital network (PHN).

The various works of Farrell have emphasized these results [14] and the use of a nonparametric frontier.

*Technical efficiency model with input orientation (1).*

$$\begin{cases} ET_{phn} = \max \lambda \\ \text{Subject to} \\ \sum_i v_i b_{u'i} \geq b_{u'phn} \quad (m = 1, \ldots, U') \\ \sum_i v_i a_{ui} \leq \lambda_{ep} a_{uphn} \quad (n = 1, \ldots, U) \\ \sum_i v_i = 1. \quad (v_i \geq 0) \end{cases} \quad (1)$$

The method determines the efficiency score to be assigned to each entity by solving the linear programming problem based on the input orientation and constant and variable returns to scale assumptions.

There are
a = 3 inputs and b = 4 outputs for public hospital networks (PHNs);
phn = 77 DMUs for 2017–2020 (DMU represents a province or prefecture public hospital network (PHN));
A as the input matrix (u × phn);
B as the output matrix (u' × phn);
z: a vector of constants (n × 1) that measures the weights used to measure the location of a DMU. In the above problem,
λ is a scalar ranging from 1 to ∞.

**Productivity gain: Malmquist index.**

The Malmquist index (MI) facilitates the analysis of total production efficiency over two or more periods, providing a measure of total factor productivity that complements the DEA method (Malmquist 1953). The objective is to analyse overall production efficiency across multiple periods; for this purpose, we apply the DEA method in combination with the Malmquist production index (MPI), which is a preferred tool for panel data analysis. Malmquist laid the foundation for this index, and the MI framework was introduced in the DEA literature by (Caves et al. 1982) (Färe et al. 1994). (Malmquist 1953) introduced the MI framework in the DEA literature. The MI always compares two adjacent periods (Health 2017, 2018, 2019, 2020).

The MI always compares two adjacent periods (Ray 2004). $T_o^t(u^t, a^t)$ and $T_o^{t+1}(u^{t+1}, a^{t+1})$ are intraperiod distance functions.

Malmquist Production Index in Periods t and t+1.

$$M_o^t = \frac{T_o^{t+1}(u^{t+1}, a^{t+1})}{T_o^t(u^t, a^t)} \quad M_o^{t+1} = \frac{T_o^{t+1}(u^{t+1}, a^{t+1})}{T_o^{t+1}(u^t, a^t)} \quad (2)$$

The *results are* decomposed by equation (2) *for* two adjacent periods of *total factor productivity* (TFP) (t and t+1).

$$M_o(u^{t+1}, a^{t+1}, u^t, a^t) = \left[\left(\frac{T_o^t(u^{t+1},a^{t+1})}{D_o^t(u^t,a^t)}\right)\left(\frac{T_o^{t+1}(u^{t+1},a^{t+1})}{T_o^{t+1}(u^t,a^t)}\right)\right]^{1/2} \quad (3)$$

$$M_o(u^{t+1}, a^{t+1}, u^t, a^t) = \frac{\frac{T_o^{t+1}(u^{t+1},a^{t+1})}{D_o^t(u^t,a^t)}}{\left[\left(\frac{T_o^t(u^{t+1},a^{t+1})}{T_o^{t+1}(u^{t+1},a^{t+1})}\right)\left(\frac{T_o^t(u^t,a^t)}{T_o^{t+1}(u^t,a^t)}\right)\right]^{1/2}} \quad (4)$$

$$MPI = TFP = \frac{T_{ov}^{t+1}(u^{t+1},a^{t+1})}{T_{ov}^{t+1}(u^t,a^t)} \left[\frac{T(u^{t+1},a^{t+1})/T_{oc}^{t+1}(u^{t+1},a^{t+1})}{T_{ov}^{t+1}(u^t,a^t)/T_{oc}^{t+1}(u^t,a^t)} \times \frac{T_{ov}^t(u^{t+1},a^{t+1})/T_{oc}^t(u^{t+1},a^{t+1})}{T_{ov}^t(u^t,a^t)/T_{oc}^t(u^t,a^t)}\right]^{1/2} \quad (5)$$

If the MPI is greater than 1, the change in TFP is positive, and vice versa.

**Modelling Peer**

One advantage of the DEA method is its ability to provide benchmarking by calculating peers for each inefficient healthcare facility closest to the efficiency frontier through peer modelling (Coelli 1996). Efficient facilities, known as benchmarks or decision-making units (DMUs), can be improved by analysing best practices developed by their efficient counterparts (Coelli 1996).

Facility B is the only DMU on the efficiency frontier based on constant returns to scale and is the model used to follow all inefficient healthcare facilities. Figure 1 illustrates peers with varying returns to scale. Three hospitals, namely, A, B, and E, are located on the efficiency frontier, indicating their optimal efficiency. On the other hand, hospitals C and D are considered inefficient. Hospital C is represented by hospitals B and E, both of which are positioned on the VRSTE-I (input orientation) frontier segment. This segment represents the projection of point C onto the VRS frontier. Similarly, hospital D has two reference peers, A and B, which are positioned on the DVRSTE-I borderline, representing the projection of point D onto the VRS frontier (Huguenin 2013).

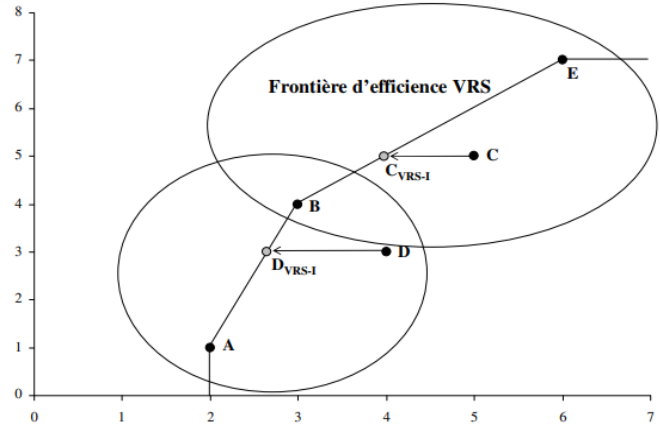

Figure 1 Example of a reference peer (**Coelli 1996; Huguenin 2013**)

## 4 Results

### 4.1 3.1 Descriptive statistics

Table 2 and Figure 2 below summarize the essential descriptive information for paramedical medicine, physicians, hospitals, functional capacity, hospitalization days, admissions, and surgical interventions.

When the central tendency measurements are examined, the median provides a robust representation of the dataset that accounts for probable outliers. Medians of 149, 50, and 2 give a deeper understanding than mean values in the context of paramedical, physician, and hospital education, especially when the data distribution is skewed.

Dispersion measures, such as the standard deviation, emphasize the degree of variability within each variable. Notably, the standard deviations for paramedical, physician, and hospital data are 393, 226, and 2, respectively, revealing the degree to which the data are scattered around the mean.

The range, as shown by the maximum and minimum values, depicts the range of values inside each variable. The maximum and minimum numbers for hospitalization days, for example, are 588,290 and 1,365, respectively, suggesting a broad spread in the dataset.

The coefficient of variation is a relative measure of variability that offers information about the fraction of dispersion from the mean. A value of 4 for paramedical and physician and 2 for hospital represents the relative variability in relation to their means in this context.

Table 2. Descriptive statistics

|  | Paramedical | Physician | Hospital | Functional capacity | Hospitalization days | Admissions | Surgical interventions |
|---|---|---|---|---|---|---|---|
| Simple Mode | 81 | 37 | 1 | #N/A | #N/A | #N/A | #N/A |
| Median | 149 | 50 | 2 | 146 | 21676 | 6969 | 1889 |
| Standard Deviation | 393 | 226 | 2 | 403 | 95369 | 16289 | 5967 |
| Maximum | 2337 | 1470 | 8 | 2387 | 588290 | 82159 | 30868 |
| Minimum | 67 | 18 | 1 | 32 | 1365 | 591 | 117 |
| Mean | 268 | 111 | 2 | 279 | 57850 | 13800 | 3941 |
| Coefficient of Variation | 4 | 4 | 2 | 3 | 4 | 2 | 3 |

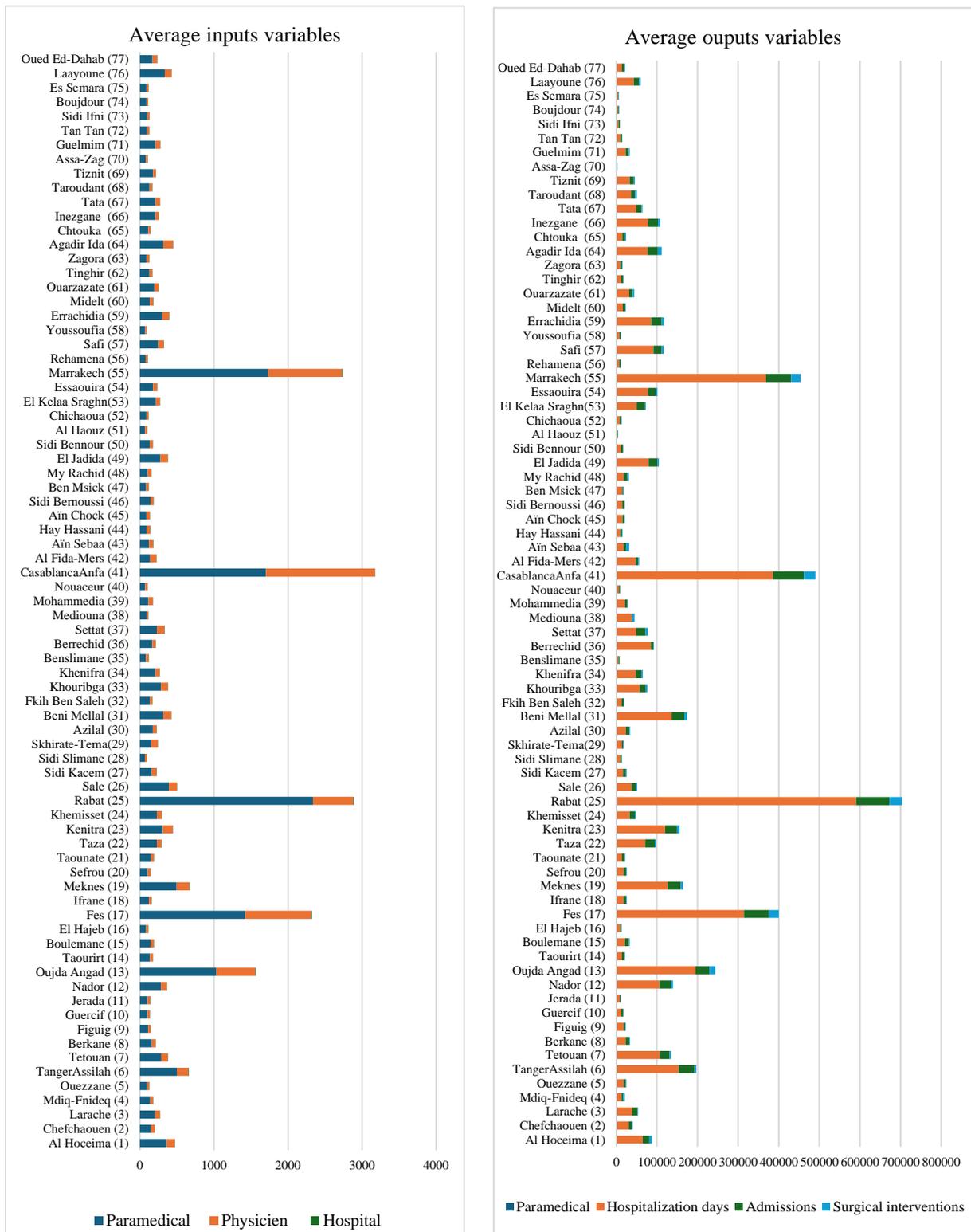

Figure 2. Average input and output variables

## 4.2 Data Envelopment Analysis

During the period of 2017–2020, the hospital network's technical efficiency scores displayed mixed performance. Under the input orientation, the average efficiency scores for the years 2017 to 2020 were as follows: Constant

Returns to Scale Assumption (CRSTE-I): 0.71; Variable Returns to Scale Assumption (VRSTE-I): 0.77; and Scale Score: 0.92.

These scores indicate the level of efficiency achieved by the hospital network in utilizing its resources during that period. However, it is important to note that without further context or information, it is challenging to provide a comprehensive analysis of these scores or their implications (Figure 3 and 4).

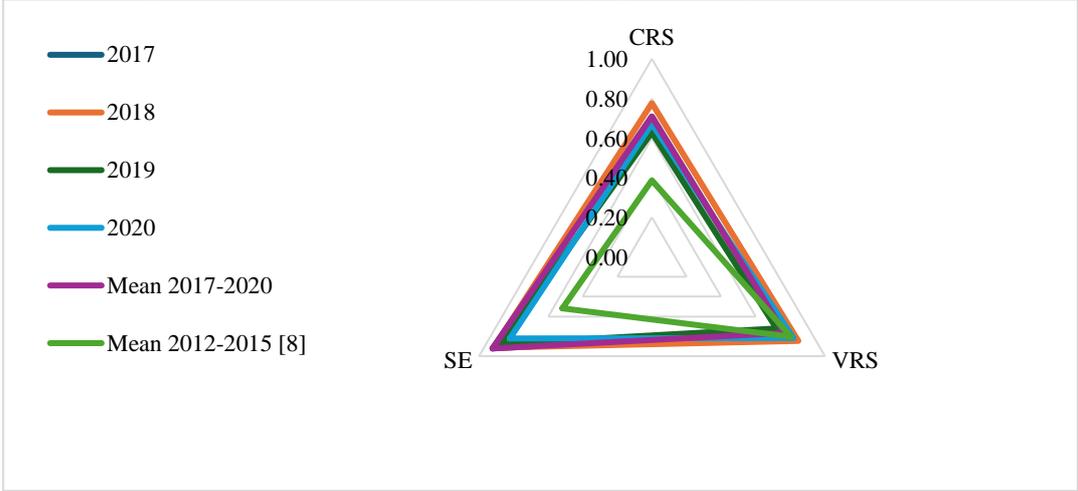

Figure 3. Mean efficiency score (PHNP/P) 2012-2015 [8] and 2017-2020

As a result, hospitals will be able to reduce their resources by an average of 29% under the CRSTE-I assumption and 27% under the VRSTE-I assumption throughout the 2017–2020 timeframe while maintaining the same level of care delivery. The scale efficiency score is 0.92, indicating that these hospitals may save 8% of their resources while maintaining the same level of care delivery in the public hospital network (Figure 3 and 4).

Furthermore, the PHNP/P network exhibited low CRSTE-I scores, with values of 0.709, 0.776, 0.629, and 0.672 in 2017, 2018, 2019, and 2020, respectively. This network might have reduced the number of inputs to 29.1% in 2017, 22.4% in 2018, 37.1% in 2019, and 32.8% in 2020 while maintaining the same output quantities (Figure 3). A comparison of the results of the study during the period of 2012-2015 [8] (Figure 3 and 4) revealed that under the Inputs orientation (I), the average score of technical efficiency for the PHNP/P network was 0.387 under the CRSTE-I assumption, indicating an efficiency of 38.7%. However, the average efficiency score under the VRSTE-I assumption reaches 80.44%. Under the input orientation approach, the hospital network can improve its input level by 61.3% under the CRSTE-I assumption but can improve by 19.58% under the VRSTE-I assumption, considering the obtained outputs.

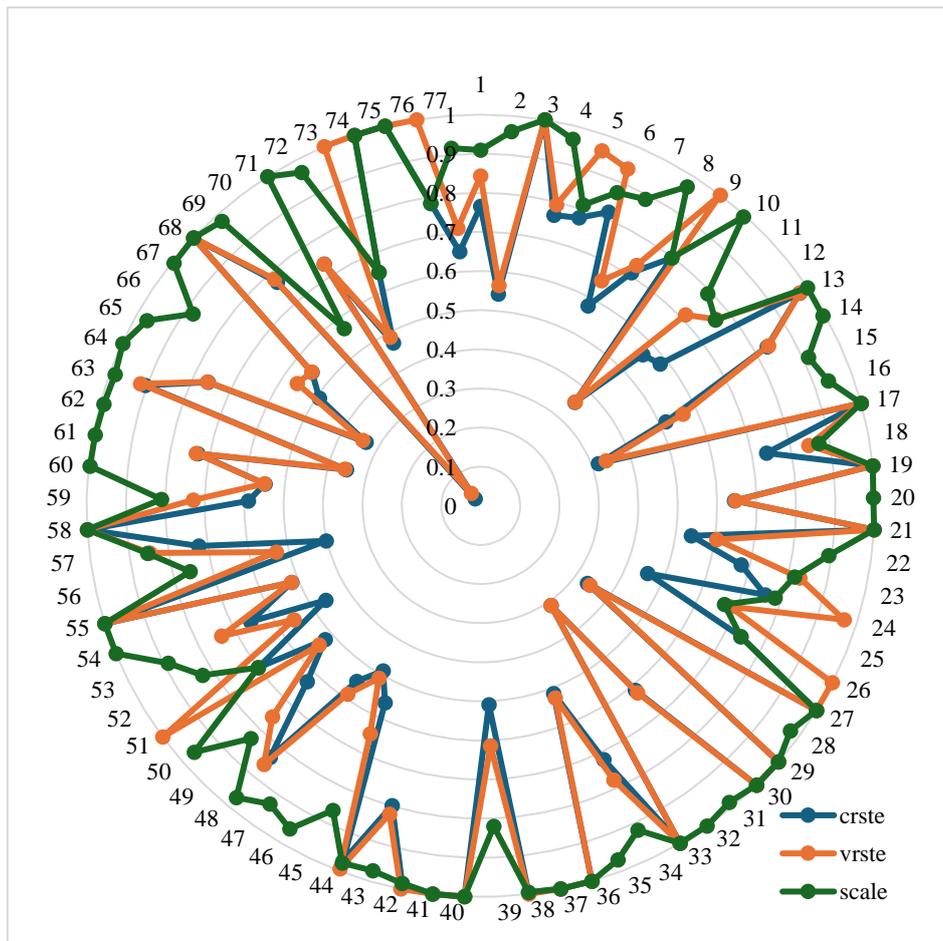

Figure 4. Average Technical Efficiency of the PHNP/P 2017–2020

### 4.3 Malmquist Index

During the period 2017–2020, the average total factor productivity (TFP) of public hospital networks decreased. The results demonstrate that 74% (57 DMUs) of hospital networks have a productivity increase of greater than one, which explains why these networks' total productivity factors have improved. In contrast, 14.4% had a Malmquist index greater than one during the study period of 2012-2015 [8] (Figure 5 and 6).

In 2019-2020, 30% of hospital networks (23 DMUs) experienced a deterioration in productivity, as indicated by an MI score less than 1. Among these hospitals, Mediouna (DMU = 38) had the lowest score of 0.191. In the preceding period, 2018-2019, 34% of hospital networks (26 DMUs) experienced a similar decline in productivity. This included Tan Tan (DMU = 72), with a score of 0.585. The lowest number of DMUs with such a decline was observed in 2017-2018, accounting for 51% (39 DMUs). Tiznit (DMU = 69) had the lowest score in this period, measuring 0.364.

We observe that the TFP (tfpch) of the PNH in Morocco was 0.980 in 2017/2018, with a 4% loss in productivity gain, owing to technological advancement (techch) with a score of 0.856 and a 14.4% deterioration. While there was a 16.3% increase in productivity gain (tfpch) in 2018/2019, this change was mostly attributable to a 49% increase in technical progress (techch).

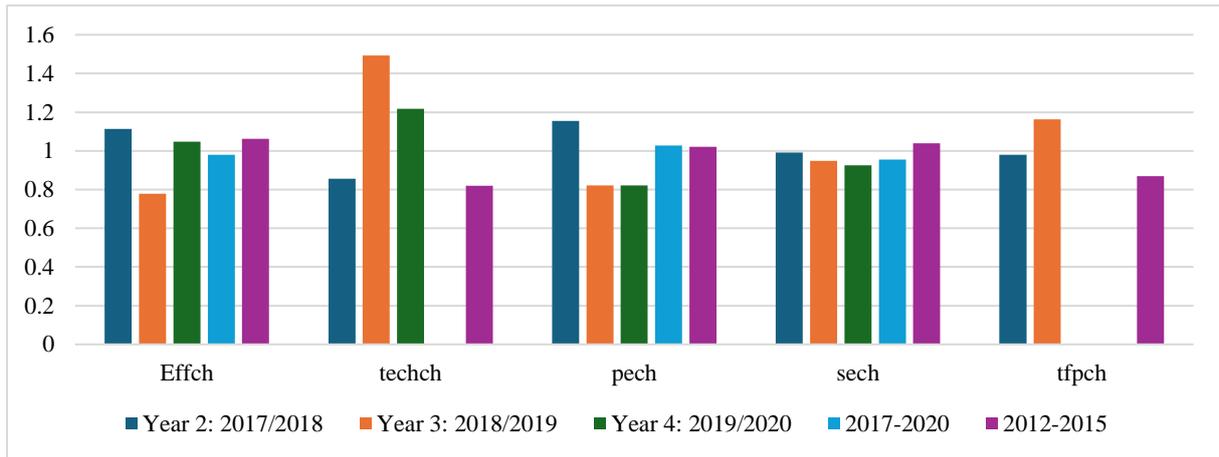

Figure 5. Malmquist Index mean PHNP/P 2012-2015 and 2017-2020

Compared to the study during 2012-2015 [8], the MI (tfpch) for the PHNP/P networks was 0.869. This decrease is attributed to the technological changes (techs) that Morocco needs to address to keep up with technological advancements and break away from traditional methods. The PHNP/P networks exhibited a technological change score of 0.819, representing an 18.1% decline (figures 5 and 6).

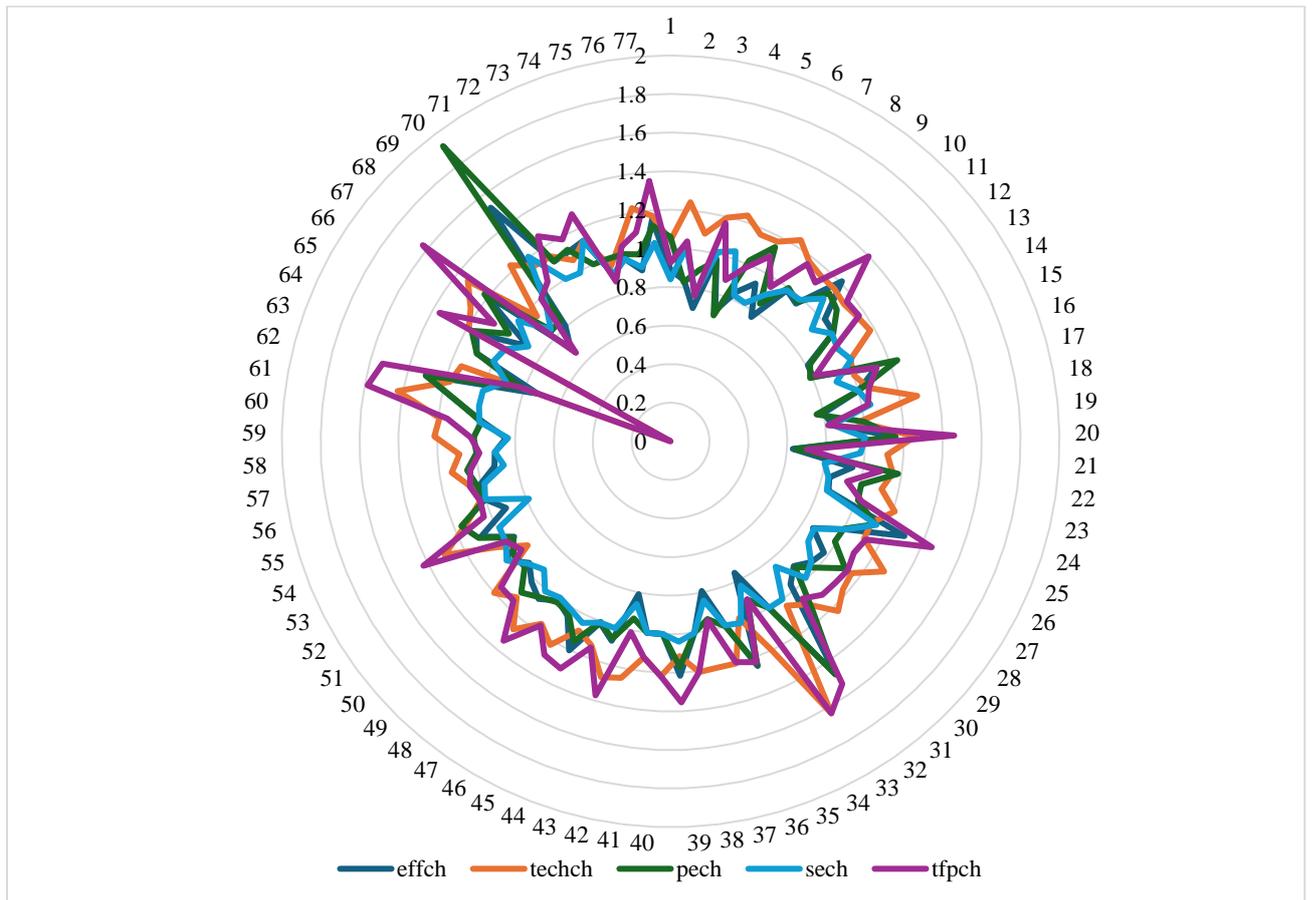

Figure 6. Average Malmquist indices for the PHNP/P 2017–2020 period

### 4.4 Peer Modelling

From the obtained results, it is evident that 72.7% of the DMUs need to emulate the most effective DMUs in 2020, whereas the lowest score is recorded in 81.8% of the 2019s. For instance, in the northern region, include Al Hoceima (DMU 1), Chefchaouen (DMU 2), Larache (DMU 3), Mdiq-Fnideq (DMU 4), Ouezzane (DMU 5), Tangier Assil (DMU 6), and Tetouan (DMU 7).

Between 2017 and 2020, the technical efficiency of the Larache DMU decreased. The efficiency score in 2017 was one, but in 2020, the Larache DMU had the lowest CRS score of 0.343, while the Mdiq-Fnideq network (DMU 4) had a higher technical efficiency score in this region, reaching 0.709 under the CRS hypothesis and 0.710 under the VRS hypothesis.

On the other hand, we observe an improvement compared to the technical efficiency study for the period 2012-2015 [8]; thus, this score was 0.117 under the CRS hypothesis.

The resources utilized by this network do not operate optimally to provide the same healthcare services. The average number of functional beds is approximately 237, the average number of hospitalization days is approximately 30,961, the average number of hospital admissions is approximately 12,340, and the average number of surgical operations is approximately 1,029.

The descriptive statistics provided by the Ministry of Health in 2020 (Health, 2020) for the PHNP/P network in Larache use various input factors to measure hospital efficiency. There were 235 nursing staff members in Larache, seventy-five physicians, and two average EHPP/P pairs. This indicates that each hospital in Larache has an average of 37.5 physicians and 115.5 nurses (across all specialties). There is also one doctor for every 6,925 people and one nurse for every 2,248 people.

This situation appears insufficient to meet the healthcare needs of the population in this province. In comparison, in the province of Mdiq-Fnideq, there were 195 nursing staff members, thirty-nine doctors, and two average PHNPs/P. This means that, on average, there are 19.5 doctors and 97.5 nurses (across all specialties) per hospital in the Mdiq-Fnideq cohort. Moreover, there is one doctor for every 6,257 inhabitants and one nurse for every 1,251 inhabitants.

It has been found that certain provinces lack doctors, nurses, and beds, but they are technically efficient. This means that even though resources are limited, they have managed to use them efficiently, as demonstrated by the case of the Mdiq-Fnideq province, which achieved an average score of technical efficiency of 0.709 (Health, 2020). However, the resources used by the network Larache do not seem to be optimally utilized to provide equivalent primary healthcare services. In fact, the average number of functional beds is approximately 237, the average number of hospitalization days is approximately 30,961, the average number of hospital admissions is approximately 12,340, and the average number of surgical operations is approximately 1,029. In this context, the DMU Larache should reduce inputs to produce the same quantities.

To achieve efficiency comparable to that of other provinces, the DMU Larache could reduce its resources used, taking inspiration from examples of efficiency in provinces such as the Peer of Hay Mohammadi PHNP/P Ain Sbaa (43), with a Peer Lambda Weight score of 0.069; the Peer Sale (26), with a Peer Lambda Weight score of 0.131; Khemisset (24), with a score of 0.705; and El Kelaa des Sraghna (53), with a score of 0.094, as shown in Table 3.

Table 3. Summary of Peer Weights pour PHNP/P Larache (2020)

| Projection Summary, Results for PHNP/P: 3 | | | | | | |
|---|---|---|---|---|---|---|
| Technical efficiency = 0.729, Scale efficiency = 0.470 (DRS) | | | | | | |
| Variable | Original value | Radial movement | Slack | Projected value | Listing of peers: | Peer Lambda weight |
| Output 1 | 191 | 0 | 52 | 243 | Casablanca-Anfa (41) | 0.000 |
| Output 2 | 70 | | 0. | 70. | Hay Mohammadi-Ain Sbaa (43) | 0.069 |
| Output 3 | 2 | 0 | | 3 | Sale (26) | 0.131 |
| Output 4 | 237 | 0 | 0 | 237 | Khemisset (24) | 0.705 |
| Input 1 | 30 | -8 | 0 | 22 | El Kelaa Sraghna (53) | 0.094 |
| Input 2 | 961 | -261 | 0 | 700 | | |
| Input 3 | 12 | -3. | 0 | 9 | | |

## 5   Discussion

The healthcare system is currently undergoing a transition, with many reforms and policies being adopted in recent years. However, the fundamental reason for the system's existing inadequacies may be linked not only to the Ministry of Health's administration but also to the management of public hospital networks themselves. To examine the impact of these reforms on the efficiency of public healthcare institutions, we used the DEA technique, the Malmquist index, and peer modelling to estimate efficiency and changes in productivity from 2017 to 2020. The findings showed that recent reform efforts have had a favorable impact on the efficiency of these facilities.

The results showed that the average technical efficiency scores of public hospital networks in Morocco under the input orientation during the period of 2017–2020 were greater than those during the period of 2012–2015 [8]. The public hospital network obtained a score of 0.71 under CRSTE-$_I$ in the period from 2017–2020, compared to 0.387 in the period from 2012–2015.

These results are lower than those in line with findings from other African countries; for example, Top et al. reported that 58.33% (21 DMUs) of 36 African healthcare systems were efficient, with Senegal being the most inefficient country. Tobit regression analysis revealed that the number of nurses per thousand people and Gini coefficient variables significantly impacted national healthcare system inefficiency (Top et al. 2020). Kirigia et al. (2001) reported that 30% of facilities in Kwazulu-Natal Province, South Africa, were technically efficient, while 70% were inefficient (Kirigia 2001). Kirigia et al. (2013) reported an average technical efficiency score of 90.3%, with 96.9% being variable and 93.3% being scale efficient in Eritrea (Kirigia et Asbu 2013). Tlotlego et al. reported that 16 of 21 hospitals in Botswana had inefficient runs in 2006, 2007, and 2008, with average VRSs of 70.4%, 74.2%, and 76.3%, respectively (Tlotlego et al. 2010). Ali et al. (2017) reported that 6 (50%), 5 (42%), 3 (25%), 3 (25%), 4 (33%), and 3 (25%) of the hospitals were technically inefficient under VRS, while 9 (75%), 9 (75%), 7 (58%), 7 (58%), 7 (58%) and 8 (67%) of the hospitals were scale inefficient between 2007/08 and 2012/13, respectively. On average, the Malmquist total factor productivity (MTFP) of hospitals decreased by 3.6% over the panel period (Ali et al. 2017).

During the COVID-19 crisis, the Ministry of Health's commitment resulted in a considerable increase in healthcare expenditure, from 8% in 2018 to 12% in 2020, resulting in improved technical efficiency and reinforced hospital facilities and healthcare personnel. While the functioning of healthcare facilities (PHNPs/Ps) has focused on preventive rather than curative treatment, these facilities have been ineffective at treating health concerns since the country's independence. This situation is also crucial for addressing the issues posed by the COVID-19 epidemic. Despite a slight improvement in healthcare provision during the study period of 2017–2020, primarily due to increased infrastructure; financial and material resources; and medical, paramedical, and administrative personnel, the overall performance of primary healthcare and healthcare facilities remains poor.

The central issue does not revolve around the number of infrastructures, financing, universal healthcare coverage, or the quantity of human resources. Instead, it pertains to equitable access to care, the quality of healthcare provision, the selection of managerial practices, and the lack of mechanisms and a culture of control and audit in healthcare facility management. This article serves as a foundational study to stimulate scientific debate on the advantages and disadvantages of hospital care providers. The study concluded that the efficiency of healthcare facilities influences access to care through quality services and improved patient management. Hospital networks that provide healthcare services attract a larger number of patients from inefficient provinces.

## 6    Conclusion and Recommendations

This study analysed the technical efficiency of the PHNP/P in Morocco using the DEA method, the Malmquist index, and peer modelling. A minority of the provinces approach the efficiency frontier, while most do not effectively utilize their inputs in healthcare provision. This is due to the unstable and discontinuous implementation of reforms. Inefficient provinces should continuously monitor their personnel activities and periodically publish their performance for public review to identify weaknesses and enable necessary corrections.

The study's limitations include not methodologically explaining the factors leading to inefficiencies, which is difficult to understand due to the lack of information on the technical functioning of healthcare establishments. The COVID-19 pandemic has highlighted the need for policymakers to prioritize infrastructure, motivate personnel, and implement sector-specific policies. The study recommends monitoring service delivery effectiveness, identifying underutilized inputs, reducing the average length of stay, and increasing the physician-to-staff ratio.

The study's conclusions are influenced by organizational aspects, health determinants, and the need to reconcile organizational constraints with consumer preferences. Further research is needed to expand the list of input and output variables from an objectivist perspective.

### Abbreviations

CRSTE: Constant return to scale technical; DEA : Data Envelopment Analysis, DEAP: Data Envelopment Analysis Programming, DMU: Decision Making Unit, IM: Index Malmquist, Effch : Technical Efficiency Change, ET : Technical Efficiency, pech : Pure Change, PHN : Public Hospital Networks, PHNP/P : Provincial or Prefectural Public Hospital Networks, SE : Scale Efficiency, sech : Scale Change, Techch : Technology Change, TFP: Totals Factors Production, tfpch : Totals Factors Productivity and VRSTE: Variable return to scale technical efficiency.

**Ethics approval and consent to participate**
Not applicable

**Consent for publication**
Not applicable

**Availability of data and material**
Data from Health Ministry Moroccan

**Competing interests**
The author declare that they have no competing interests.

**Funding**
Not applicable

**Authors' contributions**
YE and MM involved in the literature review, data analysis, interpretation of the results, and drafting of the manuscript. This author read and approved the final manuscript.

**Acknowledgements**
We are immensely grateful to the Health Ministry Moroccan for their cooperation in data collection and making available the Annual Health Hospital Activity Report 2017, 2018, 2019, and 2020. We are also thankful to Ibn Tofail University for facilitating coordination for data collection for the study.

**Author details**
[1] Ibn Tofail University, Economics and Management Faculty (FEG), National School of Commerce and Management (ENCG), Laboratory: Research Laboratory in Management Sciences of Organizations, Kenitra, Morocco